\documentclass[aps, prb, preprint, superscriptaddress]{revtex4}
\usepackage{graphicx}
\usepackage{dcolumn}
\usepackage{bm}
\usepackage{natbib}
\usepackage{upgreek}
\usepackage{amsmath}
\usepackage{amssymb}

\newcommand*{\Sm}{SmFe$_3$(BO$_3$)$_4$}
\newcommand*{\vect}[1]{\mathbf{#1}}

\begin{document}

\title{Giant gigahertz optical activity in multiferroic ferroborate}

\author{A. M. Kuzmenko}
\affiliation{Prokhorov General Physics Institute, Russian Academy of
Sciences, 119991 Moscow, Russia}
\author{A. Shuvaev}
\author{V. Dziom}
\author{Anna Pimenov}
\author{M. Schiebl}
\affiliation{Institute of Solid State Physics, Vienna University of
Technology, A-1040 Vienna, Austria}
\author{A. A. Mukhin}
\author{V.Yu. Ivanov}
\affiliation{Prokhorov General Physics Institute, Russian Academy of
Sciences, 119991 Moscow, Russia}
\author{L. N. Bezmaternykh}
\affiliation{L.V. Kirensky Institute of Physics Siberian Branch of RAS,
660036 Krasnoyarsk, Russia}
\author{A. Pimenov}
\affiliation{Institute of Solid State Physics, Vienna University of
Technology, A-1040 Vienna, Austria}

\begin{abstract}

In contrast to
well studied multiferroic manganites with a spiral structure, the
electric polarization in multiferroic borates is induced within collinear
antiferromagnetic structure and can easily be switched by small
static fields. Because of specific symmetry conditions, static and
dynamic properties in borates are directly connected, which leads to
giant magnetoelectric and magneto\emph{di}electric effects. Here we
prove experimentally that the giant magnetodielectric effect in
samarium ferroborate \Sm~ is of intrinsic origin and is caused by an
unusually large electromagnon situated in the microwave range. This
electromagnon reveals strong optical activity exceeding 120 degrees
of polarization rotation in a millimeter thick sample.

\end{abstract}

\date{\today}


\maketitle

\section{Introduction}

Rapid progress of modern electronics requires a continuous search
for new mechanisms of controlling electric and magnetic properties
of materials. One of the promising recent developments targets
materials with the magnetoelectric effect which allows to influence
electric properties by magnetic field and magnetization by electric
voltage
\cite{fiebig_jpd_2005,ramesh_nmat_2007,eerenstein_nature_2006,tokura_science_2006}.
In view of future applications, the absolute value of
magnetoelectric coupling is of crucial importance. One newly
discovered material class with record values of the magnetoelectric
effect are rare-earth borates
\cite{vasiliev_ltp_2006,kadomtseva_ltp_2010,liang_prb_2011},
RFe$_3$(BO$_3$)$_4$ and RAl$_3$(BO$_3$)$_4$ (R = rare earth ion).
Especially in ferroborates with R = Sm, Ho colossal magnetic
field-induced changes in the dielectric constant have been observed
\cite{mukhin_jetpl_2011,chaudhury_prb_2009}  exceeding $\Delta
\varepsilon / \varepsilon \sim 300 \%$. Up to now such unusually
large changes in the dielectric constant were known to arise
because of extrinsic effects, such as domain wall motion
\cite{kagawa_prl_2009} or contact and grain boundary effects
\cite{lunkenheimer_prb_2002}. In the case of \Sm~ it has been suggested
that an intrinsic magnetoelectric excitation may be
responsible for the observed effects
\cite{mukhin_jetpl_2011,kuzmenko_jetp_2011}. Such excitations in
magnetoelectric materials are called electromagnons
\cite{pimenov_jpcm_2008,sushkov_jpcm_2008}, and they are defined as
magnetic excitations that interact with the electric component of
electromagnetic radiation.

Although the existence of certain magnetoelectric modes in \Sm~
may be expected from the basic arguments, all relevant frequency
ranges except for the microwaves could be excluded in previous
experiments. In this work we prove the existence of a
magnetoelectric excitation at gigahertz frequencies. The observation
of the electromagnon in \Sm~ becomes possible because the
eigenfrequency of the mode can be lifted to the millimeter-wave
range by the external magnetic field. Because of the strong coupling of static
and dynamic magnetoelectric properties in \Sm~ giant controlled
polarization rotation is demonstrated.

Magneto-optical effects in multiferroics represent an intensive and rapidly developing field of investigations. The examples include magnetic field-induced
dichroism in the terahertz range
\cite{pimenov_nphys_2006,kida_prb_2011}, controlled chirality \cite{bordacs_nphys_2012},
directional dichroism
\cite{kezsmarki_prl_2011,takahashi_nphys_2012,takahashi_prl_2013}.
Electric control of terahertz radiation is more difficult to realize
and it has been recently demonstrated in the millimeter wave-range \cite{shuvaev_prl_2013} and by Raman scattering \cite{rovillain_nmat_2010}.

In ferroborates RFe$_3$(BO$_3$)$_4$ the unusual strength of the
magnetoelectric coupling results from the presence of two magnetic
subsystems: iron and rare-earth~\cite{kadomtseva_ltp_2010}. Existing
data suggest that the interaction between the two subsystems increases
the magnetoelectric coupling in ferroborates by at least one order of
magnitude~\cite{zvezdin_jetpl_2006}. A distinctive feature of the
borates is that their crystal structure is non-centrosymmetric which
is in contrast to manganites with a perovskite-like structure
(RMnO$_3$). This results in different symmetry conditions for
magnetoelectric properties. In particular, in ferroborates
electric polarization is induced by the external magnetic field or by
collinear antiferromagnetic ordering of Fe-ions while in the
manganites it appears only within non-centrosymmetric (cycloidal)
antiferromagnetic ordering of Mn ions. The microscopic mechanism of
magnetoelectricity in borates is still unknown. However, it
seems to be clear that the same mechanism is responsible for static
and dynamic magnetoelectric effects. This coupling promises
desirable direct connection between static and gigahertz properties, which may open up novel applications such as new
effective ways to control millimeter-wave light with external
voltage or magnetic field. The connection described above differs considerably from that in rare-earth manganites. In manganites
the static polarization is determined by Dzyaloshinski-Moriya
coupling~\cite{katsura_prl_2007,mostovoy_prl_2006}  and the dynamic
properties are mainly governed by the Heisenberg exchange
mechanism~\cite{aguilar_prl_2009,lee_prb_2009}. This incompatibility
hampers the control of dynamic properties by static fields.

\section{Experimental}

Spectroscopic experiments in the terahertz frequency range (40~GHz
$< \nu <$ 1000 GHz) have been carried out in a Mach-Zehnder
interferometer arrangement~\cite{volkov_infrared_1985} which allows
measurements of the amplitude and the phase shift in a geometry with
controlled polarization of radiation. Theoretical transmittance
curves \cite{shuvaev_sst_2012} for various geometries were
calculated from the susceptibilities using Fresnel optical equations
for the complex transmission coefficient and within the Berreman
formalism \cite{berreman_josa_1972}. Details of the terahertz data processing are given in Appendix \ref{app:B}. The experiments in external
magnetic fields up to 7~T have been performed in a superconducting
split-coil magnet with polypropylene windows. Static dielectric
measurements have been done using commercial impedance analyzer
equipped with a superconducting magnet. Large single crystals of \Sm, with typical dimensions of $\sim 1$
cm, have been grown by crystallization from the melt on seed
crystals.

\section{Samarium Ferroborate}

\begin{figure}[tbp]
\begin{center}
\includegraphics[width=0.95\linewidth, clip]{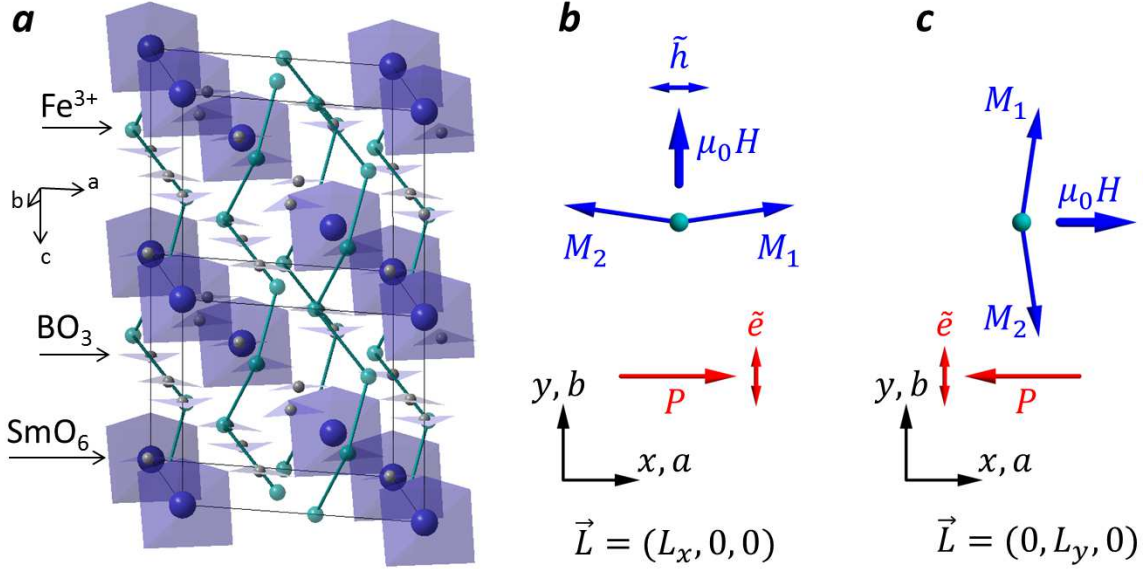}
\end{center}
\caption{\emph{Magnetic structure of \Sm.} \textbf{a} - Basic
structural elements in \Sm. \textbf{b} - Orientation of Fe$^{3+}$
magnetic moments and the excitation conditions of the
magnetoelectric mode (electromagnon) in external magnetic fields
parallel to the crystallographic $b(y)$-axis. \textbf{c} - changes
in the magnetic structure and excitation conditions for $\mu_0 H \|
a(x)$-axis.} \label{fig1}
\end{figure}

\Sm~ contains two interacting localized magnetic subsystems given by
Sm$^{3+}$ and Fe$^{3+}$ ions. The iron subsystem orders
antiferromagnetically below $T_N=34~ K$ with an easy-plane magnetic
structure oriented perpendicularly to the trigonal c-axis. Although
the Sm$^{3+}$ moments play an important role in the magnetoelectric
properties of \Sm, they probably do not order up to the lowest
temperatures. The crystallographic structure of \Sm~ is shown in
Fig.~\ref{fig1}\textbf{a}. \Sm~ has a non-centrosymmetric trigonal structure with R32 space group \cite{vasiliev_ltp_2006}.

Static electric polarization in multiferroic ferroborates can be
explained by symmetry arguments and by taking into account that
Fe$^{3+}$ moments are oriented antiferromagnetically within the
crystallographic
$ab$-plane~\cite{zvezdin_jetpl_2005,zvezdin_jetpl_2006}. Within the
topic of the present work the term governing the ferroelectric
polarization along the $a$ and $b$-axis (or $x$ and $y$-axis, see
Fig.~\ref{fig1}\textbf{a}) is of basic importance. For the R32 space group of
borates this term is given by
\begin{equation}\label{p}
P_x \sim L_x^2-L_y^2 ,\quad  P_y \sim -2L_xL_y \ .
\end{equation}
Here $L=M_1-M_2$ is the antiferromagnetic vector with $M_1$ and
$M_2$ being the magnetic moments of two (antiferro-)magnetic
Fe$^{3+}$ sublattices. Full details of the symmetry analysis of the
static magnetoelectric effects in \Sm~ can be found in Refs.
\cite{zvezdin_jetpl_2005,zvezdin_jetpl_2006,popov_prb_2013}.

Simple expression Eq.~(\ref{p}) allows to understand the behavior of
static and dynamic properties in external magnetic fields.  As shown
in Fig.~\ref{fig1}\textbf{b}, static magnetic field along the $y$-axis
stabilizes the magnetic configuration with $L_y=0$ and $L_x\neq 0$.
In agreement with Eq.~(\ref{p}) in this case the static polarization
is oriented parallel to the a-axis. For magnetic fields along the
$x$-axis and above the spin flop value $L_x=0$ and $L_y\neq 0$ which
leads to antiparallel orientation of electric polarization with
respect to the a-axis (Fig.~\ref{fig1}\textbf{c}).

Details of the model analysis of the magnetoelectric modes in \Sm~
are given in Appendix \ref{app:A}. In brief, the magnetic and
electric excitation channels are connected because of direct coupling of
the antiferromagnetism and ferroelectricity. The low-frequency
magnetoelectric mode of interest corresponds to oscillations of
antiferromagnetic moment $L$ in the easy $ab$-plane. It can be
excited either by an ac electric field with $e \bot P$ or by an ac
magnetic field $h \bot M$. Here $P$ is the static electric
polarization and $M\| \mu_0 H$ is a weak field-induced ferromagnetic
moment with $M \bot L$. Because electric polarization is directly
coupled to the antiferromagnetic order, it becomes possible to
excite the spin oscillations not only by an ac magnetic field but by
an alternating electric field as well. The excitation conditions of
the electromagnon strongly differ for two magnetic configurations
given in Figs.~\ref{fig1}\textbf{b,c}. In particular, the excitation
conditions of the electromagnon for the configuration in
Fig.~\ref{fig1}\textbf{c} are given by $e\|b$ or $h\|b$. Therefore, for an
ab-cut of \Sm~ crystal we can selectively excite either electric
($e\|b$) or magnetic ($h\|b$) component of the electromagnon simply
by rotating the polarization of the incident radiation.

The excitation conditions of the electromagnon change substantially
if the static magnetic field is parallel to the $b$-axis as shown in
Fig.~\ref{fig1}\textbf{b}. In this case the external field stabilizes the
configuration with magnetic moments parallel to the $a$-axis and
with $L_y=0$. In agreement with Eq.~(\ref{p}), the static electric
polarization is parallel to the $a$-axis. The excitation
conditions of the electromagnon now changes to $e\|b$ and $h\|a$.
Therefore, with this configuration of the magnetic moments and for
an $ab$-cut sample the electric and magnetic excitation channels are
either simultaneously active (for the polarization $e\|b,h\|a$) or
they are both silent (for $e\|a,h\|b$).
\begin{figure}[tbp]
\begin{center}
\includegraphics[angle=270, width=0.95\linewidth, clip]{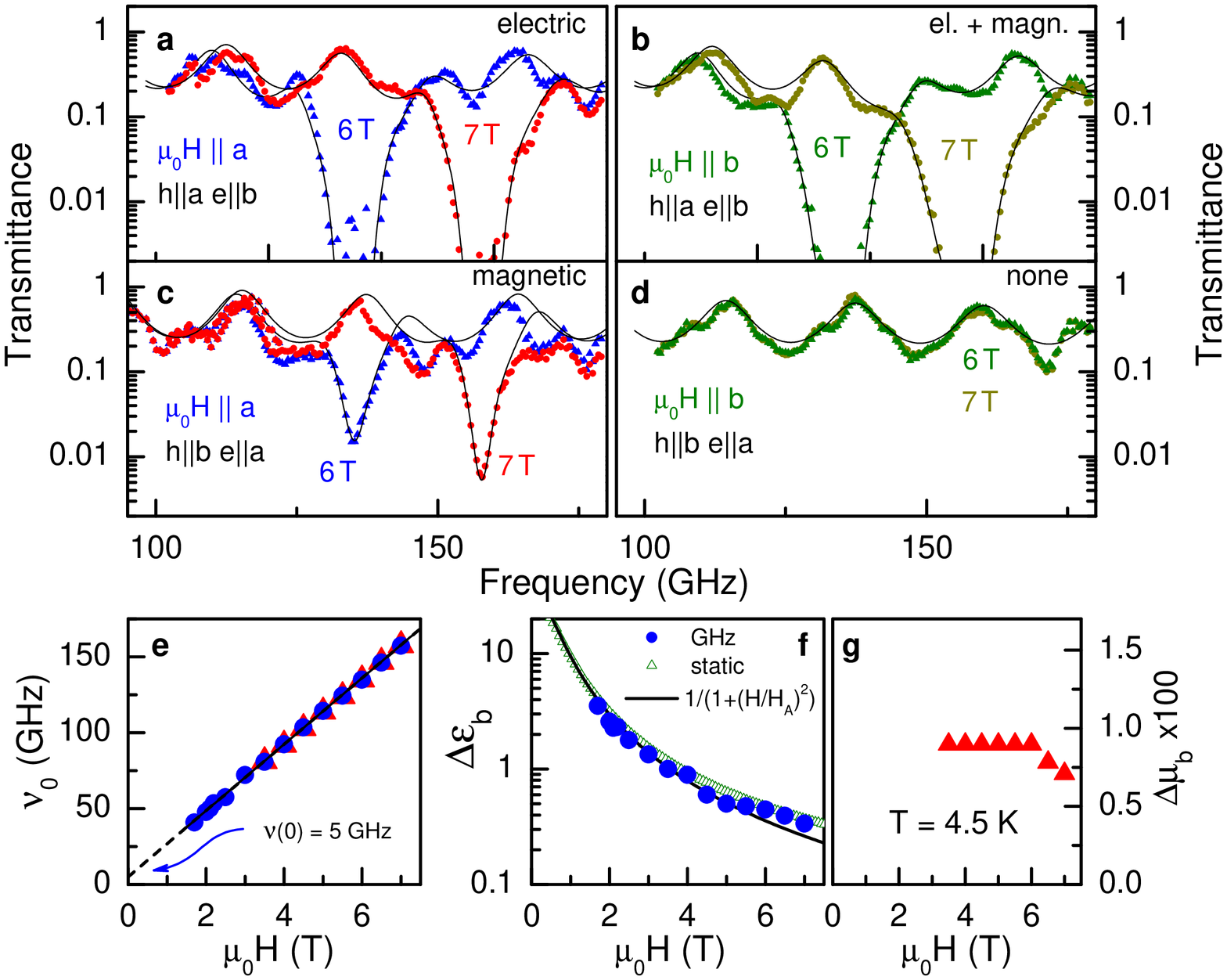}
\end{center}
\caption{\emph{Selective excitation of electromagnon in \Sm.}
\textbf{a-d} - Transmittance spectra for two orientations of
external magnetic field, $\mu_0 H \|a $ (\textbf{a,c}) and  $\mu_0 H
\|b $ (\textbf{b,d}). Different excitation conditions allow to
selectively probe electric (\textbf{a}), magnetic (\textbf{c}),
magnetoelectric (\textbf{b}), or silent (\textbf{d}) geometry.
\textbf{e} - Magnetic field dependence of the electromagnon
frequency. Circles - $\mu_0 H \|a , h\|a, e\|b $, triangles - $\mu_0 H \|a , h\|b, e\|a $, straight line is a linear fit to the resonance frequency
with $\nu_0(0 T)= 5$~GHz. \textbf{f,g} - Field dependence of the
electric and magnetic contribution of the electromagnon. Circles -
electric contribution, red full triangles - magnetic contribution,
green open triangles - static permittivity, solid line - model
calculations.} \label{figem}
\end{figure}

\section{Results and Discussion}

Typical transmittance spectra of \Sm~ in the frequency range of our
spectrometer are shown in Figs.~ 2\textbf{a-d}. These results
support the symmetry arguments given above and they can be
consistently described by an electromagnon revealing an electric dipole
moment parallel to the $b$-axis and with the magnetic dipole moment
which can be switched between $h\|a$ and $h\|b$ depending on the
orientation of the static magnetic field.

The eigenfrequency of the electromagnon in \Sm{} in zero magnetic
field is determined by the weak magnetic anisotropy in the basis
$ab$-plane and is estimated as $\nu_0 \approx 5$ GHz. This frequency
increases roughly linearly with external magnetic field as
demonstrated in Fig. 2\textbf{e}. Strong tunability of the
electromagnon frequency in external fields makes multiferroic
ferroborates  an attractive material class for applications. As
demonstrated in Fig.~\ref{figem}\textbf{f} and in agreement with the
model given in Appendix \ref{app:A}, the dielectric
contribution of the electromagnon is initially saturated at the
static value of $\sim 35$ but decreases in external fields as
\begin{equation}\label{delta}
    \Delta \varepsilon = \Delta \varepsilon_0/[1+H^2/(2H_E H'_A)] \ .
\end{equation}
Here $H_E$ is the exchange field, $H'_A$ is the anisotropy field in
the basis plane and $\Delta \varepsilon_0 \simeq 30$ is the static
dielectric contribution.

As mentioned above, the symmetry of the R32 space group of \Sm~
allow a static electric polarization which can be rotated by
magnetic field. An immediate consequence for dynamic properties is
the existence of strong nonzero magnetoelectric susceptibility of
electromagnon in~\Sm. Experimentally, these terms in the
electrodynamic response would lead to a rotation of the polarization
plane in the frequency range of the mode. Because of the strong dielectric
contribution of the electromagnon ($\Delta \varepsilon \sim 30$)
large values of the optical activity can be expected.
\begin{figure}[tbp]
\begin{center}
\includegraphics[width=0.95\linewidth, clip]{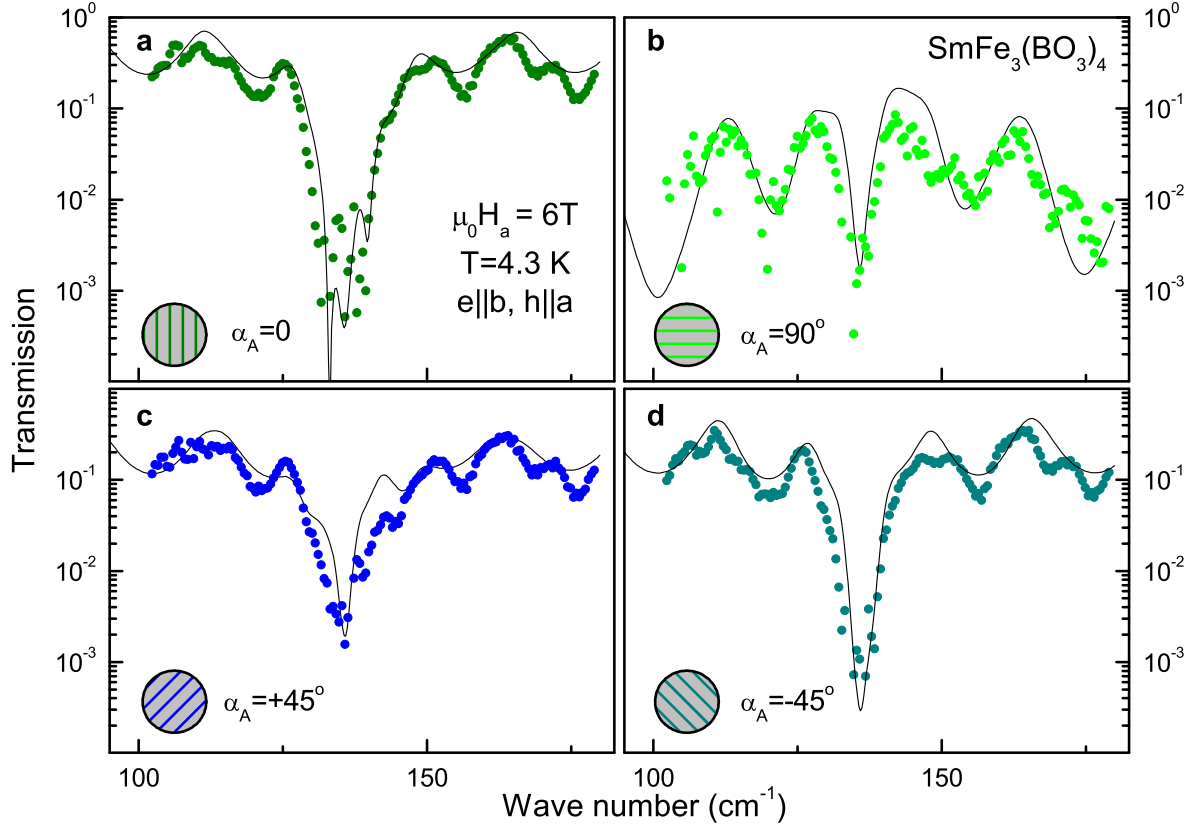}
\end{center}
\caption{\emph{Optical activity in \Sm}.
Transmittance spectra for various mutual orientation of the
polarizer and analyzer. \textbf{a} - Parallel geometry. \textbf{b} -
Crossed geometry. \textbf{c,d} - $\pm45^{\circ}$ geometry.   Symbols
- experiment, solid line - model calculations.} \label{fig45}
\end{figure}

In order to prove that the dynamic magnetoelectric susceptibility
induces optical activity in \Sm, we investigated the polarization
state of the transmitted radiation as shown in
Fig.\ref{fig45}\textbf{a,b}. In these experiments the incident beam
is linearly polarized. The transmitted power is measured for the
analyzer rotated by the angle 0$^{\circ}$, $90^{\circ}$, and
$\pm45^{\circ}$. Without optical activity the signal within
$90^{\circ}$ (Fig.~\ref{fig45}\textbf{a}) would be zero and the $\pm45^{\circ}$ spectra (Figs.~\ref{fig45}\textbf{c,d}) would
coincide, which evidently contradict the spectra in
Fig.~\ref{fig45}. The evaluation of the transmitted power at
different analyzer angle allows to fully characterize the
polarization state of the transmitted radiation without additional
measurements of the phase information. For example, the rotation
angle $\theta $ and the ellipticity $\eta$ are given by
\begin{equation}\label{theta}
    \tan(2 \theta)=
    \frac{T(+45^\circ)-T(-45^\circ)}{T(0^\circ)-T(90^\circ)} \
    ,\
    \sin(2 \eta) =
    \frac{\sqrt{4 T(0^\circ)T(90^\circ)-[T(+45^\circ)-T(-45^\circ)]^2}}{T(0^\circ)+T(90^\circ)}
    \ ,
\end{equation}
where $T(+45^\circ),T(-45^\circ),T(0^\circ)$, and $T(90^\circ)$ are
the power transmission values for analyzer angle rotated by
$+45^\circ,-45^\circ,0^\circ$, and $90^\circ$, respectively. This equation follows directly from Eq.~\ref{eq45}.
\begin{figure}[tbp]
\begin{center}
\includegraphics[angle=270, width=0.95\linewidth, clip]{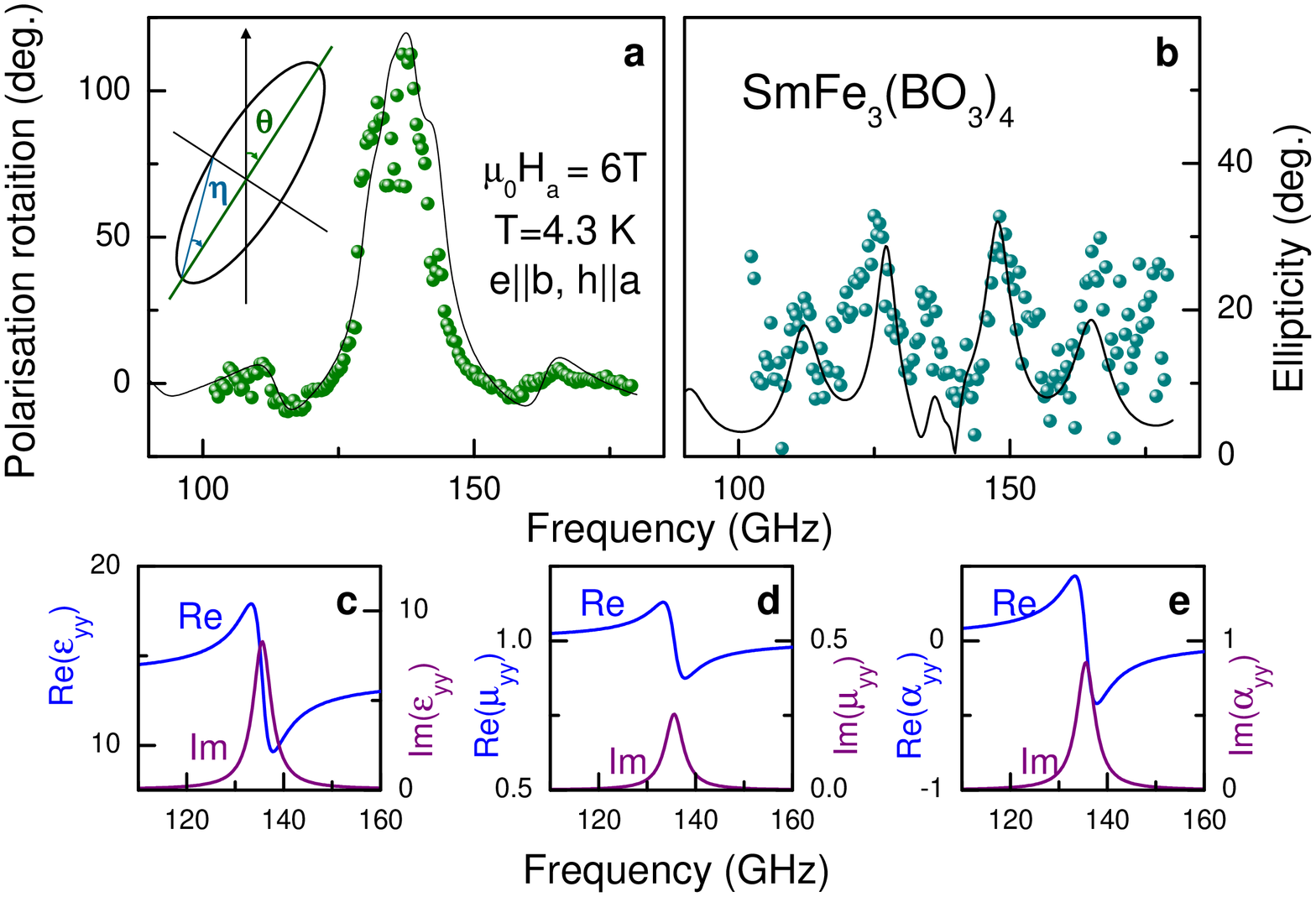}
\end{center}
\caption{\emph{Polarization rotation by electromagnon in \Sm}.
\textbf{a} - Polarization rotation ($\theta$). The inset shows the
definition of rotation angle and ellipticity. \textbf{b} -
Ellipticity ($\eta$). \textbf{c-e} - electric, magnetic, and
magnetoelectric permittivities as obtained from the model analysis
of the transmission spectra at $\mu_0 H = 6$~T. Symbols -
experiment, solid line - model calculations. } \label{figrot}
\end{figure}
The angle of the polarization rotation and the ellipticity are shown
in Fig. \ref{figrot}\textbf{a,b}. It is a remarkable result of these
experiments that a polarization rotation angle exceeding 120 degrees
is obtained for a sample with thickness of 1.7 millimeter only. We
stress that this rotation arises purely from dynamic magnetoelectric
susceptibility which is intrinsic for electromagnon in \Sm. Static
magnetic field is in this case needed solely to lift the resonance
frequency of the electromagnon into the available range of our
spectrometer. This is in contrast to the polarization rotation by
charge carriers \cite{palik_rpp_1970} or magnetic resonance
\cite{zvezdin_book} which require the Faraday geometry of the
experiment and arise because of the off-diagonal elements in electric
conductivity and magnetic permeability, respectively.

\section{Conclusions}

In conclusion, this work demonstrates experimentally that the giant magneto\emph{di}electric effect in multiferroic ferroborate \Sm~ arises as a result of a large electromagnon in the gigahertz frequency range. Based on  symmetry arguments, the electromagnon in \Sm~ reveals strong electric and magnetoelectric activity and can be controlled by external magnetic field. A polarization rotation exceeding 120 degrees is observed at gigahertz frequencies and is explained using dynamic magnetoelectric susceptibility in \Sm. Such a strong
effect allows effective control of the gigahertz radiation via
magnetoelectric effect.

\subsection*{Acknowledgements}
This work was supported by Russian Foundation for Basic Researches
(N 12-02-01261, 12-02-31461 mol), and by the Austrian Science Funds
(I815-N16, W1243).

\appendix
\section{Theory of dynamic magnetoelectric effect in
S$\rm{m}$F$\rm{e}_3$(BO$_3$)$_4$}
\label{app:A}

In order to describe the dynamic magnetic, magnetoelectric and
magnetodielectric properties of SmFe$_3$(BO$_3$)$_4$ we shall
consider the thermodynamical potential which depends on the
ferromagnetic ($\vect{m}$) and antiferromagnetic ($\vect{l}$)
vectors of the antiferromagnetically ordered Fe-subsystem, electric
polarization $\vect{P}$ and external magnetic $\vect{H}$ and
electric $\vect{E}$ fields:
\begin{equation}
\Phi(\vect{m}, \vect{l}, \vect{P}, \vect{H}, \vect{E}) =
\Phi_m(\vect{m}, \vect{l}, \vect{H}) + \Phi_{me}(\vect{m}, \vect{l},
\vect{P}) + \Phi_e(\vect{P}, \vect{E}) \ . \label{free_energy}
\end{equation}
The first term in Eq.~(\ref{free_energy}) represents the magnetic
part in the antiferromagnetically ordered state with $l \gg m$ and
$\vect{l} \bot \vect{m}$, and it is given by
\begin{equation}
\Phi_m(\vect{m}, \vect{l}, \vect{H}) = \frac{1}{2} A \vect{m}^2 -
M_0 \vect{m} \vect{H} + \Phi_A(\vect{l}),
\label{magnetic_energy}
\end{equation}
where the first and second terms are the exchange and Zeeman energy,
and the third term is the anisotropy energy
\begin{equation*}
\Phi_A(\vect{l}) = \frac{1}{2} K_{eff} l_z^2 + \frac{1}{12} K_6
\left[ (l_x + i l_y)^2 + (l_x - i l_y)^2 \right] - \frac{1}{2}
K_{1u} \left(l_x^2 - l_y^2 \right) - K_{2u} l_x l_y \ .
\label{anisotropy_energy}
\end{equation*}
The uniaxial anisotropy stabilizes the magnetic moments within the
basis $ab$ plane ($K_{eff}
> 0$) in which the anisotropy is determined by the hexagonal
crystallographic anisotropy ($K_6$) and magnetoelastic anisotropy
$K_{1u} \sim \sigma_{xx} - \sigma_{yy}$, $K_{2u} \sim \sigma_{xy}$
induced by internal stress (compression/elongation $\sigma_{xx} -
\sigma_{yy}$ and shift $\sigma_{xy}$) in the crystallographic $ab$
plane.

In Eq.~(\ref{free_energy}) the magnetoelectric energy
$\Phi_{me}(\vect{m}, \vect{l}, \vect{P})$ relevant for the present
analysis and related to the orientation of the Fe$^{3+}$ spins in
the basis $ab$ plane can be written
as~\cite{zvezdin_jetpl_2005,popov_prb_2013}:
\begin{equation}
\Phi_{me}(\vect{m}, \vect{l}, \vect{P}) = -c_2 P_x \left(l_x^2 -
l_y^2 \right) + 2 c_2 P_y l_x l_y + \mathellipsis \ .
\label{me_energy}
\end{equation}
The electric part $\Phi_e(\vect{P}, \vect{E})$ of the
thermodynamical potential  Eq.~(\ref{free_energy}) is determined by
the expression
\begin{equation}
\Phi_e(\vect{P}, \vect{E}) = \left(P_x^2 + P_y^2 \right) /
\left(2 \chi_{\perp}^e \right) + P_z^2 / \left(2 \chi_{\|}^e \right) -
\vect{P} \vect{E},
\label{electric_energy}
\end{equation}
where $\chi_{\|}^e$ and $\chi_{\perp}^e$ are the crystal lattice
parts of (di)electric susceptibility along and perpendicular to the
$c$-axis, respectively. For the sake of simplicity the contribution
of the Sm subsystem is not explicitly shown in
Eqs.~(\ref{free_energy})-(\ref{me_energy}) but it is assumed that
the corresponding parameters ($A$, $M_0$, $K_{eff}$, $c_2$,
$\mathellipsis$) are renormalized due to Sm-Fe exchange
interaction~\cite{zvezdin_jetpl_2005,popov_prb_2013}.

By minimizing the thermodynamic potential $\Phi$ in
Eq.~(\ref{free_energy}) in $\vect{P}$ one obtains the equilibrium
positions of the electric polarization:
\begin{equation*}
P_x = P_0 \left(l_x^2 - l_y^2 \right) + \chi_{\perp}^e E_x, \quad
P_y = - 2 P_0 l_x l_y + \chi_{\perp}^e E_y, \quad P_z = \chi_{\|}^e
E_z, \label{polarization}
\end{equation*}
where $P_0 = c_2 \chi_{\perp}^e$ determines the maximal spontaneous
polarization in the basis plane in a single-domain state, which is
induced by the antiferromagnetic Fe ordering. Substituting
$\vect{P}$ into Eq.~(\ref{free_energy}) one can obtain the
Landau--Lifshits equations for the dynamic magnetic variables
$\vect{m}$ and $\vect{l}$:
\begin{equation}
\left(M_0 / \gamma_{Fe} \right) \dot{\vect{m}} = \vect{m} \times \vect{\Phi}_m
+ \vect{l} \times \vect{\Phi}_l, \quad
\left(M_0 / \gamma_{Fe} \right) \dot{\vect{l}} = \vect{m} \times \vect{\Phi}_l
+ \vect{l} \times \vect{\Phi}_m,
\label{ll_equations}
\end{equation}
where $\vect{\Phi}_m = \partial \Phi / \partial \vect{m}$,
$\vect{\Phi}_l = \partial \Phi / \partial \vect{l}$ and $\gamma_{Fe}
= g_{Fe} \mu_B / \hbar$ is the gyromagnetic ratio for Fe-ions. The
linearization and solution of these equations with respect to small
oscillations of $\vect{m}$ and $\vect{l}$ in the easy plane state
$\vect{l}_0 \| \vect{b}$ stabilized by the magnetic field $\vect{H}
\| \vect{a}$ allows to derive the magnetic and electric response to
the alternating fields $\vect{e}$ and $\vect{h}$:
\begin{equation*}
\begin{array}{c}
\Delta \vect{m} = \chi^m \vect{h} + \chi^{me} \vect{e}, \\
\Delta \vect{p} = \chi^{em} \vect{h} + \chi^e \vect{e}. \\
\end{array}
\label{constitutive_equations}
\end{equation*}
Here magnetic $\chi^m$, magnetoelectric $\chi^{me}$, $\chi^{em}$ and
dielectric $\chi^e$ susceptibilities are given by
\begin{equation*}
\begin{array}{cc}
\hat{\chi}^m(\omega) = \left( \begin{array}{ccc}
\chi_{xx}^m & 0 & 0 \\
0 & \chi_{yy}^m & \chi_{yz}^m \\
0 & \chi_{zy}^m & \chi_{zz}^m \\
\end{array} \right) & \hat{\chi}^{me}(\omega) = \left( \begin{array}{ccc}
\chi_{xx}^{me} & 0 & 0 \\
0 & \chi_{yy}^{me} & 0 \\
0 & \chi_{zy}^{me} & 0 \\
\end{array} \right) \\[3em]
\hat{\chi}^{em}(\omega) = \left( \begin{array}{ccc}
\chi_{xx}^{em} & 0 & 0 \\
0 & \chi_{yy}^{em} & \chi_{yz}^{em} \\
0 & 0 & 0 \\
\end{array} \right) & \hat{\chi}^e(\omega) = \left( \begin{array}{ccc}
\chi_{xx}^e & 0 & 0 \\
0 & \chi_{yy}^e & 0 \\
0 & 0 & \chi_{zz}^e \\
\end{array} \right). \\
\end{array}
\label{susceptibilities}
\end{equation*}
where the individual terms are obtained as:
\begin{eqnarray*}
   \chi_{xx}^m &=& \chi_{\perp} L_{AF}(\omega)  \\
   \chi_{zz}^m &=& \chi_{\perp} L_F(\omega)  \\
   \chi_{yy}^m &=& \rho^2 \chi_{\perp}L_F(\omega)  \\
   \chi_{zy}^m &=& (-i \omega / \omega_F)\sqrt{\chi_{yy}^m \chi_{zz}^m} = (-i \omega / \omega_F) \rho \chi_{\perp}L_F(\omega) \\
   \chi_{yy}^e &=& \chi_{\perp}^e + \chi_{rot}^e L_F(\omega)  \\
   \chi_{xx,zz}^e &=& \chi_{\perp,\|}^e  \\
   \chi_{xx}^{me} &=& \chi_{xx}^{em} \approx 0  \\
   \chi_{yy}^{me} &=& \chi_{yy}^{em} = \rho \eta \sqrt{\chi_{\perp} \chi_{rot}^e} L_F(\omega)  \\
   \chi_{zy}^{me} &=& -\chi_{yz}^{em} = (i \omega / \omega_F) \eta \sqrt{\chi_{\perp} \chi_{rot}^e} L_F(\omega)
\end{eqnarray*}
Here $\chi_{\perp} = M_0^2 / A \equiv M_0 / (2 H_E)$ is the
transverse magnetic susceptibility, $M_0$ and $H_E$ are the
magnetization of the saturated antiferromagnetic sublattices of
Fe$^{3+}$ ions and Fe-Fe exchange field, respectively.
$\chi_{rot}^e$ is the electric susceptibility along the $b$-axis due
to the rotation of the spins in the basis plane. The magnetic field
dependence of $\chi_{rot}^e$ is given by
\begin{equation*}
    \chi_{rot}^e = \frac{\chi_{0rot}^e}{1 + H^2 / (2 H_A' H_E)}
\end{equation*}
where $\chi_{0rot}^e = (2 P_0)^2 / K_A'$ is the value of
$\chi_{rot}^e$ at $H = 0$ and $K_A' = \partial^2 \Phi_A / \partial
\varphi^2 \left|_{\varphi = \pm \pi / 2} \right.$ is the effective
anisotropy energy in the basis plane for the orientation $\vect{l}
\| \vect{b}$ $(\varphi = \pm \pi / 2)$~\cite{kuzmenko_jetp_2011}
(see also Ref.~\cite{note1}).

The functions
\begin{equation*}
   L_{F,AF}(\omega) = \omega_{F,AF}^2 / \left(\omega_{F,AF}^2 -
\omega^2 + i \omega \Delta \omega_{F,AF} \right)
\end{equation*}
determine the frequency dispersion of the electrodynamic response
near the resonance frequencies of the quasi-ferromagnetic (in-plane)
mode
\begin{equation*}
  \omega_F^2 = \gamma^2 \left(H^2 + 2 H_A'H_E \right)
\end{equation*}
and the quasi-antiferromagnetic (out-of-plane) mode
\begin{equation*}
\omega_{AF}^2 = 2 \gamma^2 H_A H_E \ ,
\end{equation*}
where $H_A' = K_A' / M_0$ and $H_A = K_A / M_0$ are the
corresponding anisotropy fields, $\Delta \omega_{F,AF}$ are the
line-widths of the modes which are determined by dissipation terms
omitted in Eq.~(\ref{ll_equations}).

The factor $\rho(H) = H / \sqrt{H^2 + 2 H_A' H_E}$ reflects the
changes of the magnetic structure with increasing magnetic field and
becomes unity in the fields exceeding 5-10~kOe, the factor $\eta =
(V^+ - V^-) / (V^+ + V^-)$ takes into account possible existence of
structural twins with opposite chirality with concentrations
$V^{\pm}$ and with opposite contributions to the electric
polarization.

The derivation of the magnetoelectric response given above has been
performed assuming that the resonance frequencies of the rare-earth
(Sm) ions determined by the exchange (R-Fe) splitting of its ground
doublet are higher than the AFMR frequencies of Fe-subsystem. This
condition is fully satisfied for the low frequency
quasi-ferromagnetic mode $\omega_F$ where the giant magnetoelectric
activity is observed. On the contrary, the frequency of
quasi-antiferromagnetic mode $\omega_{AF}$ is comparable to that of
the corresponding Sm-mode and, therefore, a coupling of Fe and Sm
magnetic oscillation cannot be neglected~\cite{kuzmenko_jetpl_2011}.
However, due to high frequencies of these modes the dynamic
magnetoelectric effect is weak and has not been observed.

\section{Data processing}
\label{app:B}

The light propagating along the $z$ direction can be characterized
by the tangential components of electric ($E_x$, $E_y$) and magnetic
($H_x$, $H_y$) fields. We write these components in the form of a 4D
vector $\vect{V}$:
\[
\vect{V} = \left( \begin{array}{c}
E_x \\
E_y \\
H_x \\
H_y \\
\end{array} \right).
\]
The interconnection between vectors $\vect{V}_1$ and $\vect{V}_2$,
corresponding to different points in space separated by a distance
$d$, is given by $\vect{V}_1 = M(d) \vect{V}_2$. Here, $M(d)$ is a
4x4 transfer matrix. The susceptibility tensor for SmFe(BO$_3$)$_4$
has the form:
\begin{equation}
\left( \begin{array}{c}
D_x \\
D_y \\
D_z \\
B_x \\
B_y \\
B_z \\
\end{array} \right) = \left( \begin{array}{cccccc}
\varepsilon_{xx} & 0 & 0 &
0 & 0 & 0 \\
0 & \varepsilon_{yy} & 0 &
0 & \alpha_{yy} & \alpha_{yz} \\
0 & 0 & \varepsilon_{zz} &
0 & 0 & 0 \\
0 & 0 & 0 &
\mu_{xx} & 0 & 0 \\
0 & \alpha_{yy} & 0 &
0 & \mu_{yy} & \mu_{yz} \\
0 & -\alpha_{yz} & 0 &
0 & \mu_{zy} & \mu_{zz} \\
\end{array} \right) \left( \begin{array}{c}
E_x \\
E_y \\
E_z \\
H_x \\
H_y \\
H_z \\
\end{array} \right). \label{eqmat}
\end{equation}

The total transfer matrix $M$ is calculated following Berreman's
method~\cite{berreman_josa_1972}. In the case of normal incidence the
electromagnetic field inside the sample has the form of a plane wave
$\exp(i (k_z z - \omega t))$. The value of $k_z$ depends on the
properties of the sample and is determined solving the Maxwell
equations within the sample. In a first step the normal field
components $E_z$ and $H_z$ are expressed in terms of four tangential
components using Maxwell's equations together with the constitutive
relations Eq.~(\ref{eqmat}) above. The remaining four equations in
four variables $E_x, E_y, H_x, H_y$ can be represented in the form
of an eigenvalue problem with 4x4 matrix. The eigenvalues give four
possible values of $k_z$ and correspond to four waves propagating
inside the sample: two polarizations in positive $z$ direction and
two polarizations in the negative direction. The polarizations of
these waves are given by the corresponding eigenvectors of the
solution. The choice of the tangential field components simplify the
calculations because $E_x, E_y, H_x, H_y$ are continuous at the
sample-air boundary. The described procedure is only slightly
modified in the case of oblique incidence.

As a next step, the transfer matrix $M(d)$, which relates vectors
$\vect{V}$ in the air (vacuum) on both sides of the sample, is
written as
\begin{equation*}
M(d) = W K(d) W^{-1} \ .
\end{equation*}
The matrix $W$  is composed of four eigenvectors of the solution
above and it transfers the tangential field components into the
waves which propagate inside the sample. The diagonal matrix $K_{jj}
= \exp(i k_z^{(j)} d)$ is constructed out of four eigenvalues
$k_z^{(j)}$, $j = 1, \ldots, 4$ and it describes the propagation of
the electromagnetic eigenmodes along the $z$-direction.

To illustrate the procedure with a simple example, we describe an
isotropic dielectric medium with simple constitutive equations
$\vect{D} = \varepsilon \vect{E}$ and $\vect{B} = \mu \vect{H}$. In
the case of normal incidence the eigenvalue problem can be written as:
\[
\frac{c k_z}{\omega} \vect{V} = \left( \begin{array}{cccc}
0 & 0 & 0 & \mu \\
0 & 0 & -\mu & 0 \\
0 & -\varepsilon & 0 & 0 \\
\varepsilon & 0 & 0 & 0 \\
\end{array} \right) \vect{V}.
\]
The matrices $W$ and $K(d)$ are given by:
\[
W = \left( \begin{array}{cccc}
Z & Z & 0 & 0 \\
0 & 0 & Z & Z \\
0 & 0 & -1 & 1 \\
1 & -1 & 0 & 0 \\
\end{array} \right), \quad K(d) = \left( \begin{array}{cccc}
e^{i k d} & 0 & 0 & 0 \\
0 & e^{-i k d} & 0 & 0 \\
0 & 0 & e^{i k d} & 0 \\
0 & 0 & 0 & e^{-i k d} \\
\end{array} \right).
\]
Here, $Z=\sqrt{\mu / \varepsilon}$ and $k = \sqrt{\mu \varepsilon}
\, \omega / c$. The resulting transfer matrix $M(d) = W K(d) W^{-1}$
takes the form:
\[
M(d) = \left( \begin{array}{cccc}
\cos(kd) & 0 & 0 & i Z \sin(kd) \\
0 & \cos(kd) & -i Z \sin(kd) & 0 \\
0 & -i Z^{-1} \sin(kd) & \cos(kd) & 0 \\
i Z^{-1} \sin(kd) & 0 & 0 & \cos(kd) \\
\end{array} \right).
\]
In order to calculate the complex transmission and reflection
coefficients it is more convenient to change the basis. In the new
basis, the first component of the vector $\vect{V}$ is the amplitude
of the linearly polarized wave ($E_x$) propagating in positive
direction, the second - of the wave with the same polarization
propagating in negative direction and the third and the fourth
components - of two waves with another linear polarization ($E_y$).
The propagation matrix in the new basis is $M' = V^{-1} M V$, with
the transformation matrix given by:
\[
V = \left(\begin{array}{cccc}
1 & 1 & 0 & 0 \\
0 & 0 & 1 & 1 \\
0 & 0 & -1 & 1 \\
1 & -1 & 0 & 0 \\
\end{array} \right).
\]
The complex transmission and reflection coefficients for the
linearly polarized incident radiation can now be found from the
following system of equations:
\[
\left(\begin{array}{c}
t_{\perp} \\
0 \\
t_{\|} \\
0 \\
\end{array} \right) = M' \left(\begin{array}{c}
0 \\
r_{\perp} \\
1 \\
r_{\|} \\
\end{array} \right).
\]

Here the $t_{\|}$ and $t_{\perp}$ denote the complex transmittance
amplitudes within parallel and crossed polarizers, respectively; the
same conventions for reflectance are given by $r_{\|}$ and
$r_{\perp}$. In the simple example above the transfer matrix
contains four relevant elements only:
\[
M' = \left( \begin{array}{cccc} c_d + i \frac{Z + Z^{-1}}{2} s_d &
i \frac{Z^{-1} - Z}{2} s_d & 0 & 0 \\
i \frac{Z - Z^{-1}}{2} s_d &
c_d - i \frac{Z + Z^{-1}}{2} s_d & 0 & 0 \\
0 & 0 & c_d + i \frac{Z + Z^{-1}}{2} s_d &
i \frac{Z^{-1} - Z}{2} s_d \\
0 & 0 & i \frac{Z - Z^{-1}}{2} s_d &
c_d - i \frac{Z + Z^{-1}}{2} s_d \\
\end{array} \right),
\]
where $c_d = \cos(kd)$ and $s_d = \sin(kd)$. The complex
transmission coefficient in this case is well known and can be
written explicitly as:

\begin{equation*}
t_{\|} = \left(\cos(kd) - i \frac{Z + Z^{-1}}{2}
\sin(kd)\right)^{-1} \ .
\end{equation*}

In particular case when the measurements are performed with the
analyzer rotated by $\pm 45^{\circ}$ to the incident radiation, the
corresponding complex transmission coefficients $t_{\pm 45}$ are
related to transmission in parallel $t_{\|}$ and crossed $t_{\perp}$
geometries as:
\begin{equation}
\left( \begin{array}{c} t_{+45} \\ t_{-45} \\ \end{array} \right) =
\frac{1}{\sqrt{2}} \left( \begin{array}{cc}
1 & 1 \\ 1 & -1 \\
\end{array} \right) \left( \begin{array}{c} t_{\|} \\ t_{\perp} \\
\end{array} \right). \label{eq45}
\end{equation}

The polarization rotation $\theta$ and the ellipticity $\eta$ are
obtained from the transmission data using \cite{palik_rpp_1970}:
\begin{eqnarray*}
&& \tan(2\theta)=2\Re(\chi)/(1-|\chi|^2)\ , \\
&& \sin(2\eta)=2\Im(\chi)/(1+|\chi|^2)\ .
\end{eqnarray*}

Here $\chi=t_{\perp}/t_{\|}$ and the definitions of $\theta + i \eta$ are
shown schematically in Fig. \ref{figrot}\textbf{a}.

\bibliographystyle{unsrt}
\bibliography{literature}

\begin{thebibliography}{10}
\bibitem{note1}
The value of the in-plane anisotropy can be either positive or negative
  depending on the interrelation between the hexagonal crystallographic
  anisotropy and the magnetoelastic anisotropy induced by internal stress.
  Moreover, its effective value and the easy axis direction in the basis plane
  can be distributed due to inhomogeneous internal stress. In this case the
  corresponding rotational electric susceptibility should be determined by
  averaging over a whole crystal. However, in magnetic fields exceeding the
  in-plane spin-flop field $\sqrt{2 H_A' H_E} \sim$~5-10~kOe when spins are
  aligned perpendicular to an external magnetic field, the behavior of the
  rotational electric susceptibility is determined basically by the
  field-induced anisotropy in the basis plane $\chi_{rot}^e \sim (2 P_0)^2 /
  \left(\chi_{\perp} H^2 \right)$.
\end{thebibliography}

\end{document}